\documentclass{article}
\usepackage{spconf,amsmath,graphicx}

\usepackage{enumitem}
\setlist{nosep, leftmargin=14pt}

\usepackage{mwe} %

\title{Improved Brain Age Estimation with Slice-based Set Networks}
\name{Umang Gupta$^{\star}$ \qquad Pradeep K. Lam$^{\dagger}$ \qquad Greg Ver Steeg$^{\star}$ \qquad Paul M. Thompson $^{\dagger}$}
\address{$^{\star}$ Information Sciences Institute, University of Southern California \\
    $^{\dagger}$
    Imaging Genetics Center, Mark and Mary Stevens Institute for Neuroimaging and Informatics,\\ Keck School of Medicine, University of Southern California \\
    }
\usepackage{color}
\usepackage{booktabs}
\usepackage{diagbox}
\usepackage{hyperref}
\usepackage{xcolor}
\hypersetup{
    colorlinks,
}
\graphicspath{{figures/}}
\usepackage{amsmath}
\usepackage{amsfonts}

\newcommand{\real}{\mathbb R}

\begin{document}
\maketitle

\begin{abstract}
    Deep Learning for neuroimaging data is a promising but challenging direction. The high dimensionality of 3D MRI scans makes this endeavor compute and data-intensive. Most conventional 3D neuroimaging methods use 3D-CNN-based architectures with a large number of parameters and require more time and data to train.
        Recently, 2D-slice-based models have received increasing attention as they have fewer parameters and may require fewer samples to achieve comparable performance.
        In this paper, we propose a new architecture for BrainAGE prediction. The proposed architecture works by encoding each 2D slice in an MRI with a deep 2D-CNN model. Next, it combines the information from these 2D-slice encodings using set networks or permutation invariant layers.
    Experiments on the BrainAGE prediction problem, using the UK Biobank dataset, showed that the model with the permutation invariant layers trains faster and provides better predictions compared to  other state-of-the-art approaches.
\end{abstract}
\begin{keywords}
    MRI, deep learning, brain age, machine learning, neuroimaging
\end{keywords}
\section{Introduction}\label{sec:intro}
In this work, we focus on the problem of predicting brain age from 3D MRI scans. \textit{Brain Age Gap Estimation} (BrainAGE)  from structural MRI acts as an important biomarker for assessing and diagnosing an individual's risk of neurological diseases. A brain age prediction model is estimated by training on a large dataset of MRIs of healthy subjects to predict their chronological age. Deviation of the true age from this MRI derived age is a useful biomarker for various neurological diseases~\cite{TenYearsBrainAge}.

Deep Convolutional Neural Networks (CNN) have shown tremendous improvements over traditional computer vision approaches. Directly extending computer vision successes to 3D neuroimaging data by substituting 2D-CNN with 3D-CNN is non-trivial and has received considerable attention~\cite{kleesiek2016deep,singh20203d}. A 3D-CNN has more parameters than its 2D counterpart and therefore requires more data to train robust models. However, the number of samples for any 3D neuroimaging problem is often less, with the largest datasets typically having tens of thousands rather than millions of samples. Conventional approaches that have applied deep learning to neuroimaging have been focused on designing better data augmentation techniques and designing better 3D-CNN models~\cite{peng2019accurate,COLE2017115,DINSDALE2021117401}. Other approaches have tried to use transfer learning~\cite{hon2017towards,MRISignBrainAge}; however, the scarcity of general-purpose pretrained 3D-CNN models has limited these methods to adapt 2D-CNN models, which is not ideal.

Recently,~\cite{Lam2dSliceRnn} proposed a model for the BrainAGE problem, which encodes slices along the sagittal axis using a 2D-CNN encoder and then processes this ordered sequence of slices using a recurrent model (the long short-term model, LSTM). Their approach outperforms the 3D counterpart when trained from scratch. However, it relies on fixing the ordering of the slices, and the optimal ordering is unclear. Moreover, slices that occur earlier in the sequence may not be able to influence the predictions much.

\begin{figure}
    \centering
    \includegraphics[width=0.47\textwidth]{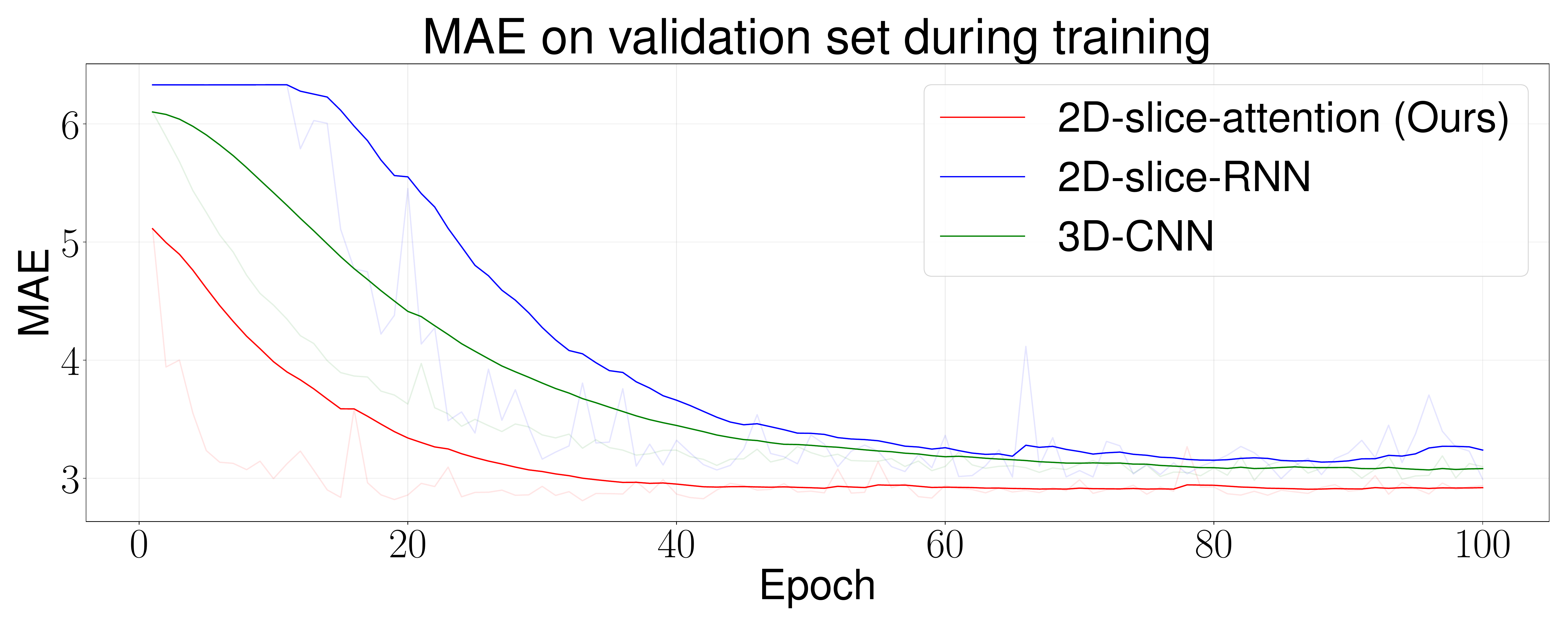}
    \caption{Smoothed training curves (original data is shown with less opacity and same color). Our proposed architecture yields better predictive performance and trains faster than the other baseline architectures (see Sec.~\ref{subsec:faster_training} for details).}
    \label{fig:training_curve}
\end{figure}

Our approach to this problem alleviates the requirement of specifying the ordering over slices by using recently proposed set networks~\cite{lee2019set,zaheer2017deep}.
Similar to~\cite{Lam2dSliceRnn}, we employ a 2D encoder that encodes each slice.
However, to combine the information in different slices, we consider the slices as a set and use a permutation invariant operation over this set. The output of the permutation invariant operation does not change if the input elements are permuted. Thus, making the output independent of the ordering in which the slices are processed.
We evaluate various permutation invariant operations, namely --- mean, max, and a general weighted average operation implemented via attention.
We evaluated the proposed models on the BrainAGE prediction problem in the publicly available UK Biobank dataset~\cite{ukbb} and show that our model trains faster and provides better prediction than the other competitive baselines (see Fig.~\ref{fig:training_curve}).

\section{Model}\label{sec:model}
\begin{figure}
    \centering
    \fbox{\includegraphics[clip, trim=1.75in 1.35in 1.95in 1.85in, width=0.45\textwidth]{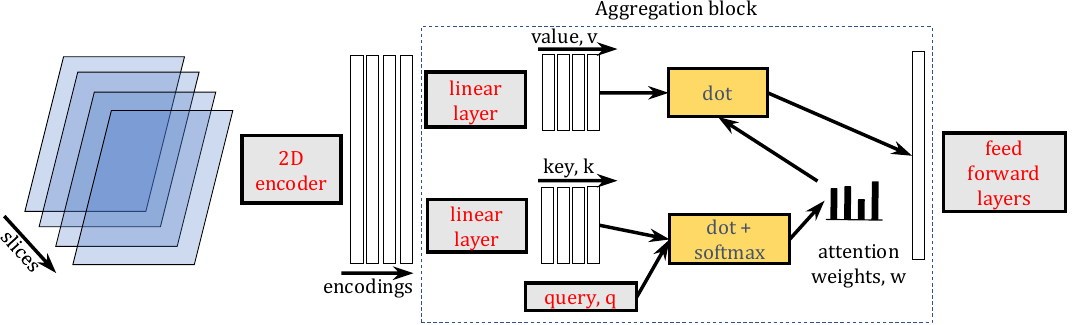}}
    \caption{Model architecture with attention-based aggregation. Gray blocks are trainable parameters, whereas yellow blocks are operations only. Each scan is considered a set of slices and transformed to a set of encodings via a 2D-CNN encoder. Attention scores are computed using these encodings and the trainable query vector. Finally, the aggregated embedding is passed through the feed-forward layers to predict the age.}
    \label{fig:architecture}
\end{figure}
Our model takes a 3D scan as input and encodes each slice using a 2D-CNN encoder. Next, it combines the slice encodings using an aggregation module (described in Sec.~\ref{subsec:aggregation}), resulting in a single embedding for the scan. Finally, we pass this embedding through the feed-forward layers to predict the brain age. The model is trained end-to-end using MSE loss. A high-level overview of our architecture is shown in Fig.~\ref{fig:architecture}.

\subsection{2D Encoder}\label{subsec:encoder}
The 2D-CNN encoder takes a single 2D slice as input and outputs a $d$-dimensional embedding for each slice. We use the same 2D encoder architecture as~\cite{Lam2dSliceRnn} - the only difference is that number of filters in the last layer is $d$, which is decided by the dimension of the output of the aggregation module, described next.

\subsection{Aggregation via Permutation Invariant Layers}\label{subsec:aggregation}
Once we have the encoding for each slice, we need to combine information across this set of slices. To this end, we use permutation invariant layers as the aggregation module;  this makes the aggregation module's output independent of the slice order.
The two most common permutation invariant operations are mean and max over the set~\cite{zaheer2017deep}; that is, we compute element-wise mean and max of all the slice encodings, respectively. Further, the mean operation can be generalized by using a weighted average of the encodings, where weights are computed using attention~\cite{lee2019set}. The attention is implemented as follows.
Let $q \in \real^{d' \times 1}$ be a trainable query vector and $r_i\in\real^{d \times 1}$ be the encoding of the $i^{th}$ slice. We first transform $r_i$ to key and value vector --- $k_i \in \real^{d'\times 1}$ and $v_i\in\real^{d{''}\times 1}$ respectively via appropriate linear layers. Next, we compute the attention scores for each encoding. If the number of slices are $p$, and $K\in \real^{d' \times p}$ be the matrix of all key vectors. The importance weights or attention is computed as  $w = \text{softmax} \left\{{q^T K}/{\sqrt{d'}}\right\}$.
Finally, we compute the weighted average of value vectors as the embedding for the scan as $\sum_i^p w_i v_i$. Multiple attention heads can be used so that the model can focus on different slices for prediction. To achieve $m$ heads, we  use $q\in \real^{d' \times m}$. %
We compute $(\sum_i^p v_i)/p$ and $\max_i^p v_i$, ignoring the query and key vectors when using mean and max operation.

For ease of reference, we name the models using mean, max, and attention operation as \textit{2D-slice-mean}, \textit{2D-slice-max}, and \textit{2D-slice-attention}, respectively.
In our experiments, we vary $d=d'=d''\in \{32, 16\}$, and vary $m \in \{1,2,4,8\}$. However, we found that the results are not very sensitive to $d$ or $m$. Therefore, we fix $d=32, m=1$ for all the models.  We use one hidden layer network with 64 activations as the feed-forward layer. We used slices along the sagittal axis; however, we find that the results do not change much if we use slices along the coronal or axial direction, as discussed in Sec.~\ref{subsec:slice_variation}.

\section{Experiment Setup}\label{sec:experiments}
\subsection{Dataset}\label{subsec:dataset}
We use the same dataset and set-up as~\cite{Lam2dSliceRnn}.
In particular, a subset of 10,446 subjects - with no psychiatric diagnosis as defined by ICD-10 criteria - was selected from 16,356 subjects in the UK Biobank dataset~\cite{ukbb}. We used the same pre-processing, and the final dimension of the images is   $91 \times 109 \times 91$.
The training, test, and validation set sizes were 7,312, 940, and 2,194, with a mean chronological age and standard deviation of  62.6 and 7.4 years.

\subsection{Baselines}\label{subsec:baseline}
\textbf{3D-CNN:}
Most conventional deep learning approaches for BrainAGE estimation use 3D-CNNs~\cite{peng2019accurate,COLE2017115}.
They adapt conventional 2D-CNN architectures to work on 3D images by replacing 2D operations with 3D operations. For instance, 2D convolutions are replaced with 3D convolutions, 2D max-pooling is replaced with 3D max-pooling, and so on. We adapt the 2D encoder mentioned in Fig.~\ref{fig:architecture} and Sec.~\ref{sec:model} to work with 3D images. Instead of using an aggregation module, we pass the encodings through another 3D convolution to produce a single node output. This architecture is the same as~\cite{peng2019accurate} but uses instance-norm instead of batch-norm after each convolutional layer due to the instability of batch-norm with smaller batch sizes.

\noindent
\textbf{2D-Slice-RNN:}
We also compare our approach to the recently proposed \textit{2D-slice-RNN} approach of~\cite{Lam2dSliceRnn}. Similar to our approach, they compute 2D encodings by taking slices along the sagittal axis. However, the sequence of encoding is aggregated by an LSTM. Their approach used fewer parameters and has been shown to outperform the corresponding 3D-CNN architecture. We use the same feature embedding size (2)  and hidden state size of LSTM (128) and apply gradient norm clipping during the training with value 1  as used in their paper.

\subsection{Training Setup}\label{subsec:training}
Each model is trained for 100 epochs with the Adam optimizer,  a weight decay of $10^{-4}$, a learning rate of $10^{-4}$, and a batch size of 8 with MSE loss. The last layer's bias is initialized with the mean age of the training set (62.68 years). We pick the best model by monitoring the performance on the validation set and report the mean absolute error (MAE) between the predicted and the true age on the test set. The code for all the experiments is publicly available at \textit{\url{https://git.io/JtazG}}.
\section{Results}\label{sec:results}
\subsection{Faster Training \& Better Predictions}\label{subsec:faster_training}

\begin{table}
    \centering
    \begin{tabular}{lcr}
        \toprule
        Method                           & MAE   & Parameters   \\
        \cmidrule(r){1-1} \cmidrule(lr){2-2} \cmidrule(r){3-3}
        3D-CNN                           & 3.017  & 2,948,801     \\
        2D-slice-RNN                     & 3.002  & 1,070,403     \\
        2D-slice-attention $(m{=}1, d{=}32)$  & 2.855  & 1,000,769     \\
        {2D-slice-mean}                  & \textbf{2.823}  & \textbf{998,625}        \\
        2D-slice-max                     & 4.213  & \textbf{998,625}     \\
        \bottomrule
    \end{tabular}
    \caption{MAE on Test set (lower is better).}
    \label{tab:mae_results}
\end{table}

Table~\ref{tab:mae_results} summarizes the mean absolute error (MAE) and the number of parameters for all the methods.
Our approach using mean and attention operations outperforms all the other methods while also being parameter efficient.
It trains faster than \textit{2D-slice-RNN} and  \textit{3D-CNN} as shown in  Fig.~\ref{fig:training_curve}.
\textit{2D-slice-RNN} suffers from the issue of having to process all the slices sequentially.
Consequently, if an important slice is towards the beginning of the sequence, it may take significant training steps to learn to propagate that information into the embedding. During the initial training phase of \textit{2D-slice-RNN}, the loss did not decrease  (see Fig.~\ref{fig:training_curve}), which supports this hypothesis. Instead, each slice encoding directly contributes to the embedding when using a permutation invariant layer, therefore receiving better gradient updates.  We see that the max operation performs poorly. We attribute this to the following --- 1) only weights corresponding to the max neuron are updated during each iteration, 2) the max operation is susceptible to outliers, so a slight change in input might lead to a large change in output. This effect can be seen in Table~\ref{tab:missing_data_result_ordered} and \ref{tab:missing_data_result_random}, where missing slices lead to less sensitivity to outliers. Thus, the performance does not degrade or even improves slightly when using max operation.
Attention generalizes the mean operation in theory; however,  its performance  is slightly worse or the same as the mean operation.

\subsection{Tolerance to Missing Slices}\label{subsec:missing}
\begin{table}
    \centering
    \begin{tabular}{lccccc}
        \toprule
        Method                   & $k{=}1$  & $k{=}2$  & $k{=}4$  & $k{=}5$  & $k{=}10$  \\
        \cmidrule(r){1-1} \cmidrule(lr){2-2} \cmidrule(lr){3-3} \cmidrule(lr){4-4} \cmidrule(lr){5-5} \cmidrule(lr){6-6}
        3D-CNN*                  & 3.01 & 3.28 & 3.77 & 3.95 & 5.21   \\
        2D-slice-RNN             & 3.00 & 3.46 & 5.65 & 6.09 & 6.98   \\
        2D-slice-RNN*            & 3.00 & 3.08 & 3.28 & 3.53 & 4.42   \\
        2D-slice-attention       & 2.85 & 2.91 & 3.02 & \textbf{3.17} & \textbf{3.77}   \\
        {2D-slice-mean}            & \textbf{2.82} & \textbf{2.87} & \textbf{3.00} &    {3.18} & \textbf{3.77}   \\
        2D-slice-max             & 4.21 & 4.10 & 3.97 & 4.11 & 4.18   \\
        \bottomrule
    \end{tabular}
    \caption{Test MAE when all but every $k^{th}$ slice is dropped, * indicates evaluation with data imputation. $k{=}1$ means no missing slices.  }
    \label{tab:missing_data_result_ordered}
\end{table}
\begin{table}
    \centering
    \begin{tabular}{lccccc}
        \toprule
        Method                   & $100\%$  & $50\%$  & $25\%$  & $20\%$  & $10\%$  \\
        \cmidrule(r){1-1} \cmidrule(lr){2-2} \cmidrule(lr){3-3} \cmidrule(lr){4-4} \cmidrule(lr){5-5} \cmidrule(lr){6-6}
        3D-CNN*                  & 3.01 & 3.51 & 4.35 & 4.60 & 5.44   \\
        2D-slice-RNN             & 3.00 & 4.96 & 6.17 & 6.42 & 6.96   \\
        2D-slice-RNN*            & 3.00 & 3.23 & 3.71 & 3.93 & 4.89   \\
        2D-slice-attention       & 2.85 & 3.02 & 3.33 & 3.45 & 4.07   \\
        {2D-slice-mean}            & \textbf{2.82} & \textbf{2.97} & \textbf{3.26} &    \textbf{3.35} & \textbf{3.91}   \\
        2D-slice-max             & 4.21 & 4.08 & 4.02 & 4.03 & 4.13   \\
        \bottomrule
    \end{tabular}
    \caption{Test MAE when slices are missing at random (averaged over 10 evaluation runs), * indicates evaluation with data imputation. Columns indicate $\%$ of slices available.}
    \label{tab:missing_data_result_random}
\end{table}

In practice, some clinical centers may use a sparser MRI acquisition (e.g., slices 5-mm apart), or scans may lack some slices due to artifacts or due to an incomplete field of view that fails to cover the entire brain. It is also of interest whether a limited slice set is sufficient, allowing reduced file transfer or faster processing, and understanding if there is redundancy in the training data.
We simulate these situations in two ways ---  1) we remove all but every $k^{th}$ slice from the scans (Table~\ref{tab:missing_data_result_ordered}); 2) we keep a fixed percentage of slices chosen at random from the scans (Table~\ref{tab:missing_data_result_random}). We do this for each scan in the test set and evaluate the models trained on complete data, i.e.,  without missing slices.
As 3D-CNN cannot be used without imputing the missing slices, we impute data by substituting the missing slices with the nearest available slice.   Our method does not depend on the ordering of the slices; therefore, it does not require imputation and can handle missing slices gracefully.
Our approach considers slices as a set rather than an ordered sequence. Therefore, It can tolerate missing elements in the set, and performance is only slightly worse than when all the data is present. Due to \textit{2D-slice-RNN}'s dependence on the ordering, it performs better with data imputation.
\subsection{Learning with less data}\label{subsec:less_data}
\begin{table}[t]
    \centering
    \begin{tabular}{lccc}
        \toprule
           Method    & $n{=}1000$   & $n{=}2500$    & $n{=}5000$   \\ %
           \cmidrule(l){1-1}  \cmidrule(lr){2-2} \cmidrule(lr){3-3} \cmidrule(lr){4-4} %
            3D-CNN   & 3.74         & 3.36          & 3.17          \\%& 3.01  \\
    2D-slice-RNN     & 3.74         & 3.43          & 3.06          \\%& 3.00  \\
2D-slice-attention   & {3.39}         & \textbf{3.12}          & \textbf{2.92}          \\%& 2.92  \\
2D-slice-mean        &\textbf{3.38}& 3.13&2.94\\
        \bottomrule
    \end{tabular}
    \caption{MAE on test set when trained with $n$  samples.}
    \label{tab:less_data}
\end{table}
Table~\ref{tab:less_data} summarizes the results when fewer training samples are available. We use a subset of $n$ samples from the training set and keep the number of updates the same as training with all the data. The performance gap between our model and the baselines is further enhanced when fewer training samples are available, suggesting the proposed model's usefulness.

\subsection{Using Slices along a Different Axis}\label{subsec:slice_variation}
\begin{table}
    \centering
    \begin{tabular}{lccc}
        \toprule
           Axis $\rightarrow$    & Sagittal  & Coronal  & Axial    \\
           \cmidrule(l){1-1}  \cmidrule(lr){2-2} \cmidrule(lr){3-3}\cmidrule(lr){4-4}
               2D-slice-attention    & 2.855    & 2.948    & 3.102         \\
               2D-slice-RNN          & 3.002& 3.266*& 3.107 \\
        \bottomrule
    \end{tabular}
    \caption{Effect of using slices along different axis, * indicates model trained with learning rate $10^{-5}$}
    \label{tab:axis_change}
\end{table}
We summarize the results of using slices along the axial and coronal axis in Table~\ref{tab:axis_change} with the \textit{2D-slice-attention} and \textit{2D-slice-RNN} model. The performance was only slightly worse than using slices along the sagittal axis.
\textit{2D-slice-RNN} was unable to learn when sliced along the coronal axis,  and we found that it was necessary to reduce the learning rate. Therefore,  we used a learning rate of $10^{-5}$.

\section{Discussion}\label{sec:discussion}
In this paper, we proposed a new 2D-slice-based architecture for BrainAGE estimation.
By considering the slices as a set and using permutation invariant layers instead of LSTM (as in~\cite{Lam2dSliceRnn}), our model combines information across slices more efficiently. It converges faster and outperforms other deep learning architectures when trained from scratch. By avoiding dependence on slice order, the proposed model is also tolerant to missing slices in the scans.

Other approaches have also employed 2D-slice-based CNN models for neuroimaging data. \cite{hon2017towards} used only slices with the highest entropy to learn the model; even so, such criteria may lead to poor outcomes, as a noisy slice can have high entropy but less information. \cite{valliani2017deep} evaluated the possibility of using pretrained 2D-residual networks for Alzheimer's disease diagnosis. \cite{islam2018brain} considers each slice as an independent sample for Alzheimer's disease diagnosis, which increases the number of samples available for training.  2D-CNNs were often chosen, as pretrained networks are widely available for 2D (but not 3D) images. \cite{MRISignBrainAge} used transfer learning with pretrained ImageNet models to predict brain age; they consider each slice as an independent sample and output the median as the prediction. When trained from scratch, this procedure yields a very high MAE (around 3.87) than the models we have discussed. 2D-slice-based approaches can be more efficient to train as they share parameters across the slices leading to fewer parameters in the model. Most of these approaches either use only a few of the slices selected via pre-processing or consider each slice as an independent sample, combining the results via ensembling. Thus, these models cannot be trained in an end-to-end fashion. Our approach combines information across all the slices using permutation invariant operations, enabling model training in an end-to-end fashion and learning to ignore any slices that are not beneficial for the task.
It is also possible to leverage transfer learning with our model. For instance, one may use a pretrained 2D encoder, and the rest of the model can be trained from scratch. An extensive comparison with transfer learning and other classical approaches is left as future work.

In Sec.~\ref{subsec:less_data}, we found that performance gaps are enhanced when fewer samples are available. This gap can be attributed to encoding slices with parameter efficient 2D-CNN rather than 3D-CNN. Even though our model encodes slices with a 2D-CNN, it is a 3D architecture when looked at end-to-end. Thus, it may provide the same expressiveness with fewer parameters. We believe that other neuroimaging prediction tasks may also benefit from this architecture.

Our proposed architecture provides improved brain age prediction for healthy subjects. To further validate the outputs as a biomarker of brain aging or neurological diagnosis~\cite{butler2020statistical, SMITH2019528}, we plan to further evaluate the model on (1) out-of-distribution samples including people with neurodegenerative diseases, and (2) data from different scanners. We also plan to test if the brain-age delta produced by our model is associated with health-related outcomes and future decline.

As this work's focus is proposing new architecture, we considered a simplified scenario,  testing the methods on one dataset without considering the scanner's effects, site, and other biases. Some recent works tackled these problems by proposing novel training objectives.
For instance,~\cite{dinsdale_harmonization} uses adversarial learning to learn a model for brain age prediction, focusing on generalization across three cohorts with different scanning protocols and age distributions.
\cite{moyer2020scanner} proposed an unsupervised method to adjust for site effects. \cite{guan_attention} used attention-based models for domain adaptation, which identify the most important brain regions to focus on. In contrast, our permutation invariant attention layer, inspired by~\cite{lee2019set}, works by identifying the most important slice.
The proposed architecture is compatible with these objectives. Future work should test how well the model generalizes in datasets with differences in scanning protocols and populations.

\section{Acknowledgments}\label{sec:Acknowledgements}

This research was supported in part by DARPA contract HR0011-2090104, and NIH grants U01AG068057 and \linebreak RF1AG051710.

\section{Compliance with Ethical Standards}
This is a study of previously collected, anonymized de-identified data available in a public repository.
Data access was approved under UK Biobank Application Number 11559.

\bibliographystyle{IEEEbib}
\bibliography{refs}
\end{document}